\documentclass{nature}

\usepackage[dvips]{graphicx} 
\usepackage{amsfonts}
\usepackage{amssymb}
\usepackage{amscd}
\usepackage{amsmath}    
\usepackage{enumerate}
\usepackage{epsfig}
\usepackage{subfigure}

\newcommand{\ket}[1]{\mbox{$\left| #1 \right\rangle$}}

\begin{document}

\title{Measurement device independent quantum key distribution over 404 km optical fibre}

\author{Hua-Lei Yin$^{1,2,\ast}$, Teng-Yun Chen$^{1,2,\ast}$, Zong-Wen Yu$^{3,4}$, Hui Liu$^{1,2}$, Li-Xing You$^5$, Yi-Heng Zhou$^{2,3}$, Si-Jing Chen$^5$, Yingqiu Mao$^{1,2}$, Ming-Qi Huang$^{1,2}$, Wei-Jun Zhang$^5$, Hao Chen$^6$, Ming Jun Li$^6$, Daniel Nolan$^6$, Fei Zhou$^7$, Xiao Jiang$^{1,2}$, Zhen Wang$^5$, Qiang Zhang$^{1,2,7,\dagger}$, Xiang-Bin Wang$^{2,3,7,\dagger}$, Jian-Wei Pan$^{1,2,\dagger}$}

\maketitle
\begin{affiliations}
\item
Hefei National Laboratory for Physical Sciences at Microscale and Department of Modern Physics, Shanghai Branch, University of Science and Technology of China, Hefei, Anhui 230026, China
\item
CAS Center for Excellence and Synergetic Innovation Center in Quantum Information and Quantum Physics, University of Science and Technology of China, Hefei, Anhui 230026, China
\item
State Key Laboratory of Low Dimensional Quantum Physics, Department of Physics, Tsinghua University,
Beijing 100084, China
\item
Data Communication Science and Technology Research Institute, Beijing 100191, China
\item
State Key Laboratory of Functional Materials for Informatics, Shanghai Institute of Microsystem
and Information Technology, Chinese Academy of Sciences, Shanghai 200050, China
\item
Corning Incorporated, Corning, New York 14831, USA
\item
Jinan Institute of Quantum Technology, Jinan, Shandong, 250101, China
\end{affiliations}
\noindent $^{\ast}$ These authors contributed equally to this work. \\
$^{\dagger}$  e-mail: qiangzh@ustc.edu.cn; xbwang@mail.tsinghua.edu.cn; pan@ustc.edu.cn

\baselineskip24pt

\maketitle
\begin{abstract}
Quantum key distribution (QKD) can provide unconditional secure communication between two distant parties\cite{BB84}.  Although the significance of QKD is undisputed, its feasibility has been questioned because of certain limitations in the practical application of real-life QKD systems.  It is a common belief the lack of perfect single-photon source and the existence of detection loss will handicap the feasibility of QKD by creating security loopholes and distance limitations\cite{Zhao:2008:quantum,lydersen2010DetBlinding}.  The measurement device independent QKD (MDIQKD)\cite{Lo:MDI:2012,Braunstein:2012} with decoy-state method\cite{H03,Wang05,LMC05} removes the security threats from both the imperfect single-photon source and the detection loss. Lengthening the distance and improving the key rate of QKD with such a superior method is thus the central issue in the practical application of QKD. Here, we report the results of MDIQKD over 404 km of ultralow-loss optical fibre and 311 km of standard optical fibre by employing an optimized four-intensity decoy-state method\cite{yzw:2015:Making}.  This record-breaking implementation of MDIQKD method not only provides a new distance record for both MDIQKD and all types of QKD systems\cite{korzh:2015:provably,shibata:2014:quantum}, more significantly, it achieves a distance that the traditional BB84 QKD would not be able to achieve with the same detection devices even with ideal single-phone sources.  For the first time, our work demonstrates that with the MDIQKD method, imperfect devices can achieve better results than what ideal sources could have achieved.  This work represents a significant step towards proving and developing a feasible long-distance QKD.
\end{abstract}

\maketitle

MDIQKD is naturally immune to all attacks against the detection system, which are believed to be the main threat to QKD. Tremendous experimental efforts have been made in
labs\cite{Rubenok:MDI:2013,Liu:MIQKDexp:2013,Silva:MDI:2013,tang:MIQKD200km:2014,Tang:MDI:2014,valivarthi:2015:measurement,pirandola:2015:high,Wang:MDI:2015,tang:2016:experimental,comandar:2016:quantum},
field tests\cite{Rubenok:MDI:2013,Tang:FieldMDIQKD:2015}, and over networks\cite{Tang:Network:2016}. So far, the longest transmission record for MDIQKD is 200 km\cite{tang:MIQKD200km:2014} in which the
key rate, 0.018 bits per second (bps), seems not to meet the requirement for practical applications. On the other hand, one of the advantages for
QKD\cite{Scarani:2009:The} is to generate fresh secure keys for instant use. This demands an appreciable final key generation in a time scale of seconds. However, prior
MDIQKD experiments show that, if statistical fluctuations are taken into consideration, one would need an ample data size to reach a considerable final key
rate\cite{Liu:MIQKDexp:2013,Rubenok:MDI:2013,Silva:MDI:2013,tang:MIQKD200km:2014,Tang:MDI:2014,valivarthi:2015:measurement,
pirandola:2015:high,Wang:MDI:2015,tang:2016:experimental,comandar:2016:quantum,Tang:FieldMDIQKD:2015,Tang:Network:2016}.
In particular, the number of total pulses at each side $N_t$ is assumed to be $10^{12}$ or even larger\cite{curty:2014:finite}, which, for a 75 MHz system\cite{tang:MIQKD200km:2014}, would
take more than 4 hours to accumulate enough data.

Several parameter optimization methods have been proposed to solve this problem\cite{Ma:2012:Statistical,Xu:2014:Protocol,Yu:2015:Statistical}. However, to fundamentally improve the key rate and
distance at an enormous scale, making only parameter optimizations is not sufficient. For long distance MDIQKD, large effects from statistical fluctuations
in estimating the phase error rate severely undermine the efficiency.
It is important to consider statistical fluctuations for different sources jointly and the worst-case estimation jointly for both yield $s_{11}$ and bit error rate $e_{11}$ of single-photon pairs\cite{yzw:2015:Making}, leading directly to the final key rate.
Here, we implement a new type of asymmetric four-intensity decoy state MDIQKD protocol\cite{yzw:2015:Making}. Each
party exploits 3 different intensities 0, $\mu_x$ and $\mu_y$ in $X$ basis, and only one intensity $\mu_z$ in $Z$ basis. The yield of single-photon pairs $s_{11}$ can be calculated by the observed gains of each two-pulse source in the $X$ basis, specifically sources $oo, ox, xo, oy,yo, xx, yy$, while source $zz$ sending out signals in the $Z$ basis is used to distill the final key.  Given the total
number of pulses, the channel loss, and the system parameters, and bounded by joint constraints of the statistical fluctuations\cite{yzw:2015:Making,Yu:2015:Statistical}, we can choose the values of $\mu_x,\mu_y,\mu_z$ and their corresponding probability distributions $p_{x}, p_{y}, p_{z}$ carefully via global optimization for all parameters to maximize the final key rate\cite{yzw:2015:Making}.

\begin{figure*}[tbh]
\centering
\resizebox{15cm}{!}{\includegraphics{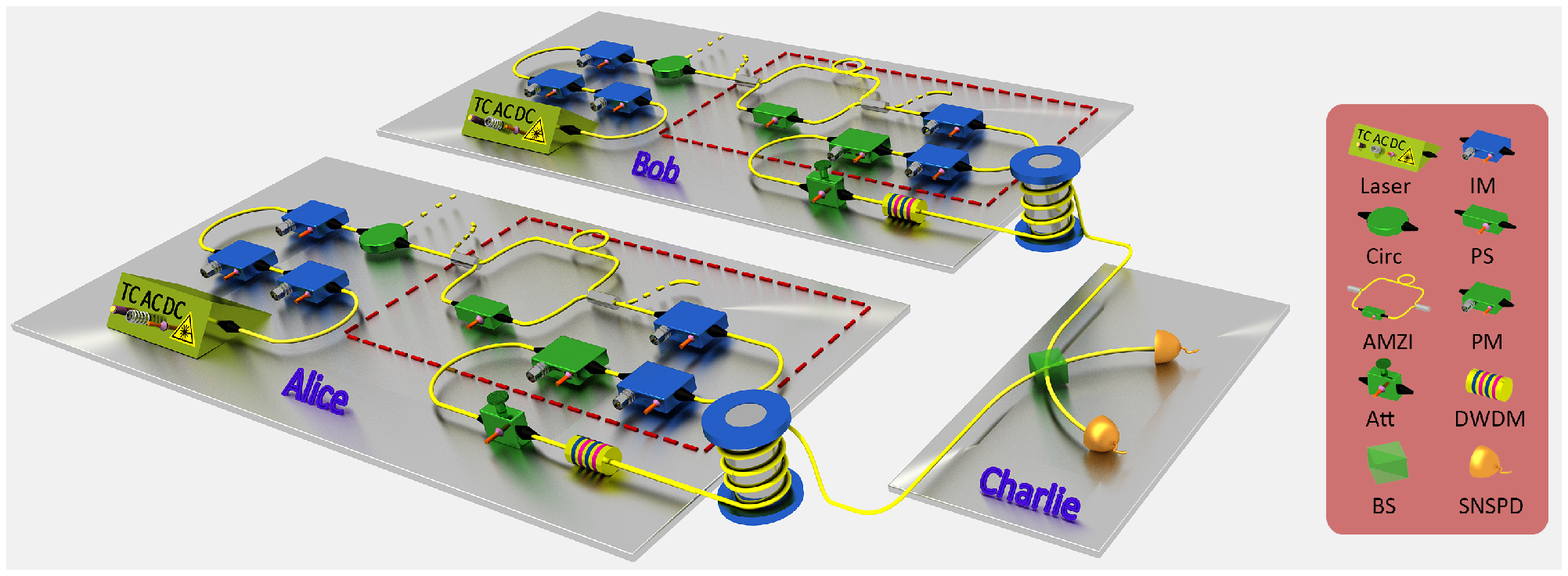}}
\caption{\textbf{Experimental setup for the MDIQKD system.} Alice's (Bob's) phase randomized weak coherent state pulses are modulated into four decoy-state intensities by two IMs.
An AMZI, two IMs and one PM are to encode time-bin phase qubits.
TC, temperature controller, AC, alternating current, DC, direct current, Circ., circulator, PS, phase shifter, IM, intensity modulator, PM, phase modulator, AMZI, asymmetrical Mach-Zehnder
interferometer, Att., attenuator, DWDM, dense wavelength division multiplexer, BS, beam splitter, SNSPD, superconducting nanowire single-photon detector.}
\label{f1}
\end{figure*}

Figure \ref{f1} schematically shows our experimental setup for MDIQKD, which consists of two identical legitimate users, Alice and Bob, and a distrustful relay, Charlie. Alice and Bob exploit
internally modulated lasers to generate phase randomized weak coherent state (WCS) optical pulses.
The pulse laser is temperature-tunable, which can be used to adjust the wavelength to implement two independent laser interference. To ensure the high visibility of two-photon interferences, an extra intensity modulator (IM) is utilized to cut off the overshoot rising edge of the optical pulse.
The full width at half maximum (FWHM) of the optimized optical pulse is 2.5 ns at a clock frequency of 75 MHz, in the wavelength of 1550.12nm. At each side, two IMs,
a phase modulator (PM), and an asymmetrical Mach-Zehnder interferometer (AMZI) are combined to form a time-bin phase qubit encoder\cite{tang:MIQKD200km:2014}. Meanwhile, two additional IMs are utilized
to add decoy states according to our optimization method\cite{yzw:2015:Making}. The intensity arrangements and probability distribution are optimized according to different transmission fibre distances.
All these modulators are controlled by random numbers independently and the corresponding radio-frequency signals come from a self-made digital to analog converter based on a field programmable gate
array.
An electrical variable optical attenuator (Att) reduces the pulse intensity down to single-photon level. Just before sending the optical pulses through the quantum channel fibre is a dense wavelength division multiplexer (DWDM) to filter
spontaneous emission noise from the laser.

Next, Alice and Bob send their pulses through optical fibres to Charlie's measurement site, respectively. An beam splitter (BS) and two superconducting nanowire single-photon detectors (SNSPDs) constitute a Bell state measurement (BSM) device. The SNSPDs operate at 2.05 K and provide detection efficiencies of 66\% and 64\% at the dark count rate of 30 counts per
second. We post-select the singlet Bell state $\ket{\Psi^-}$ when the two detectors coincide at two alternative time bins. The efficiency of time window is about 85\%, which is an optimal trade-off
between raw key rate and error rate.

To achieve a stable and enduring MDIQKD system is not a trivial task. On the one hand, our system should acquire rigorous timing and clock unification under the long transmission distance; on the other
hand, we must also solve technical complications to establish indistinguishability and calibration for the phase reference frames of Alice and Bob (see Methods). In our setup, therefore, we make use of
automatic feedback systems to calibrate the time for laser modulation, and optimize the spectrum and polarization of the two independent laser pulses from Alice and Bob.

\begin{figure*}[tbh]
\centering \resizebox{15cm}{!}{\includegraphics{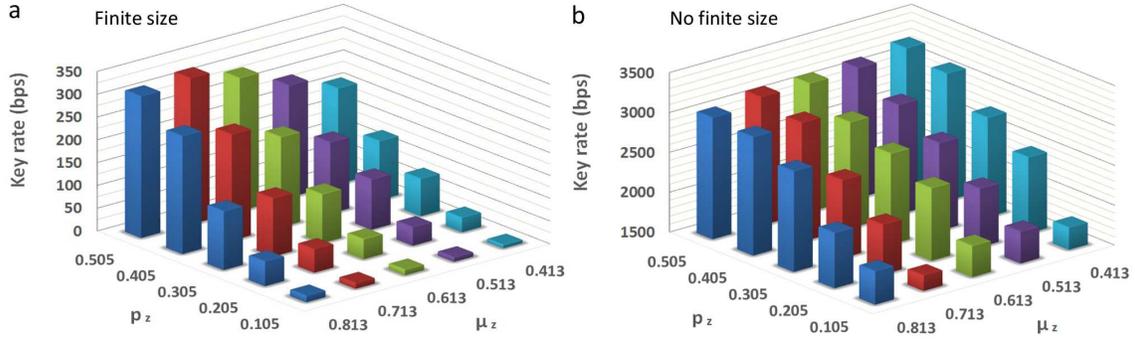}}
\caption{ \textbf{MDIQKD key rates versus intensity $\mu_{z}$ and probability $p_{z}$.} \textbf{a}, The key rates of 102 km standard fibre with a 10-minute data accumulation time. By varying the signal
state intensities $\mu_z$ and probabilities $p_z$, we achieve key rates from 7.9 to 321 bps with a failure probability of $10^{-10}$.
\textbf{b},  The key rates of the same experimental data without finite size effect. The key rates are more than 1.5 kbps for all signal state intensities $\mu_z$ and probabilities $p_z$.}
\label{f2}
\end{figure*}

One of the most outstanding properties of the 4-intensity method\cite{yzw:2015:Making} is the key rate optimization. Here we experimentally demonstrate this feature with 102 km standard optical
fibre spool. Given an error probability of $10^{-10}$, the key rate ranges from 321 to 7.9 bps with different input values of $\mu_z$ and $p_z$. Details of key rate analysis is shown in the Supplementary. This shows that given the same device, the key rate can vary
greatly from different parameters, see Fig.~ \ref{f2}~a. Note that the accumulation time and data size for each point are 10 minutes and $4.5\times10^{10}$, respectively, which are much more efficient
compared to 130 hours and $3.51\times10^{13}$ data size of the previous experiment\cite{tang:MIQKD200km:2014}, while the key rate in this work is two orders of magnitude higher than the previous one at
100 km\cite{tang:MIQKD200km:2014}. Meanwhile, as is shown in Fig.~ \ref{f2}~b, without considering finite size effect, the key rate can approach 3 kbps with the same fibre length and data accumulation
time. This reaches the level of the state-of-the-art key rate for short distance\cite{comandar:2016:quantum}.

In order to achieve longer distance, we increase the accumulation time. For example, in the case of 311 km standard fibre, we run the system for 336.6 hours and a total number of $9.09 \times 10^{13}$
pulse pairs are sent from each side in the experiment. Tens of thousands of data are collected. Setting the failure probability to $10^{-10}$, we obtained 3135 bits for our final key.

\begin{figure*}[tbh]
\centering \resizebox{15cm}{!}{\includegraphics{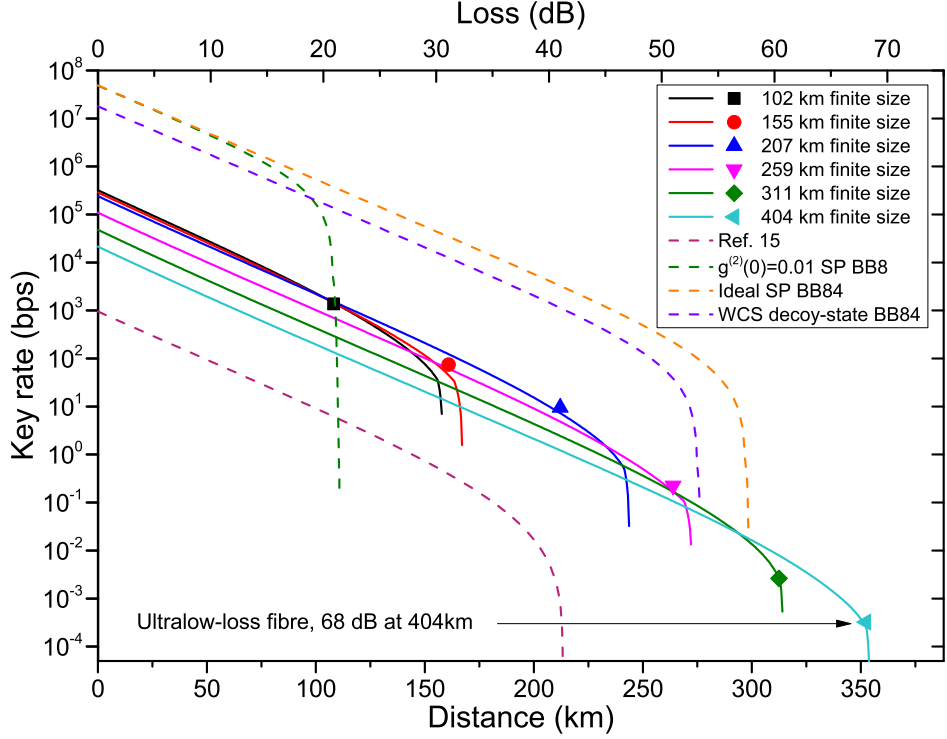}}
\caption{\textbf{Experimental results.} The experimental results (the symbols) agree well with the theoretical simulations (solid lines), with a maximum transmission distance of 404 km of ultralow-loss
optical fibre and 311 km of standard optical fibre. We also include simulations for  balanced basis passive BB84 protocol using ideal SP sources, the practical SP with $g^{(2)}(0)=0.01$ without
decoy-state method, the WCS with decoy-state method and the results of ref. 15 shown as the dotted lines for comparison. Note that although the MDIQKD produces lower key rates than BB84 protocol, it can offer further secure transmission distance.}
\label{f3}
\end{figure*}

For comparison, we implement MDIQKD at different distances of 102 km, 155 km, 207 km, 259 km, and 311 km. The optimized experimental parameters are listed in Table 1. Specifically, at the distance of
207 km, we obtain a key rate of $9.55$ bps, which is more than 500 times higher than the earlier experiment\cite{tang:MIQKD200km:2014} for the same accumulation time. Among the huge rise, about 50
times are due to the four-intensity method\cite{yzw:2015:Making,Yu:2015:Statistical} and device improvements and the increase of the efficiency of the time window can  further raise the key rate by about 10
times. The achieved key rates at various distances are shown in Fig.~\ref{f3}. The data size for each data point is listed in Table 1.
It is interesting that given the same device, at the the distance of 311 km, no secure key can be generated with the traditional passive BB84 protocol, even if we do not consider statistical fluctuations and assume that the ideal single-photon (SP) source is implemented.

Consider the passive BB84 protocol with an ideal SP source. Let $p_X$ ($p_Z$) be the probability of BS to reflect (transmit) the incident light to the measurement port of $X$ ($Z$) basis. Let $d$,
$\eta$ and $S_{\omega}$ be the dark count rate of detector, the overall efficiency and the gain of $\omega$ basis ($\omega=Z,X$), respectively.
Exploiting the linear loss model, the gain and bit error rate of $\omega$ bases is $S_{\omega}=\eta p_\omega+2d(1-d)(1-\eta p_\omega)$ and $e_{w}=d(1-d)(1-\eta p_w)/S_w$, respectively, where we assume
no alignment errors and insertion loss. In our experimental setup, the loss of 311 km standard optical fibre is $59.05$ dB while the efficiency and dark count rate of detector are $65\%$ and
$d=7.2\times 10^{-8}$, respectively. Given the fact that $0<p_X <1$ and $p_X+p_Z=1$, we find $e_{X}>7.55\%$ and $e_{X}+e_{Z}>26.25\%$. Therefore, the {\em asymptotic} key rate can be given by
$1-H(e_{X})-H(e_{Z})<0$, where $H(x)=-x\log_{2}(x)-(1-x)\log_{2}(1-x)$ is Shannon entropy.
To make it clearer, we add the theoretical curves of {\em asymptotic} key rates for balanced basis passive BB84 protocol using ideal SP, the practical SP\cite{he2013demand} with $g^{(2)}(0)=0.01$
without decoy-state method and WCS with decoy-state method in Fig. \ref{f3}.

To push the transmission limit, we test the key generation with 404 km ultralow-loss fibre (0.16 dB/km at 1550 nm) provided by Corning Incorporated. Based on the channel loss and accumulation time, we set the
optimized experimental parameters as $\mu_{z}=0.413$, $\mu_{y}=0.302$, $\mu_{x}=0.073$, $p_{z}=0.315$, $p_{y}=0.110$, and $p_{x}=0.529$. With an accumulation time of three months, we achieved 2584 bits final key with a key rate of $3.2\times10^{-4}$ bps and the result is shown in Figure 3 as well. This is by far the longest distance reported for all kinds of QKD systems.

Besides the long transmission distance, our system generates 1.38 kbits per second secure finite key at 102 km, which constitutes a strong candidate for metropolitan quantum network with untrustful
relay\cite{Tang:Network:2016}. We can further increase the system performance by increasing the system clock rate\cite{comandar:2016:quantum} and the efficiency of single photon detector\cite{marsili:2013:detecting}.

\begin{methods}

To synchronize the entire MDIQKD system and precisely overlap the two pulses from Alice and Bob at Charlie's site so as to achieve perfect interference, we improve the time calibration system and data
collection system\cite{tang:MIQKD200km:2014}. A crystal oscillator circuit exploited by Charlie generates 500 KHz electric signals, which is used for the synchronization signals of the entire systems. Charlie
sends two independent synchronization laser pulses (SLP) at 1570 nm modulated by the 500 KHz electric signals to Alice and Bob through the extra classical channel fibre, respectively. Photoelectric
detectors and phase-locked loop used by Alice (Bob) detect the SLP and regenerate  a system clock frequency of 75 MHz by frequency multiplication, respectively. Charlie controls synchronization lasers
to make Alice and Bob alternatively send specific signal optical pulses to him through quantum channel. Charlie records the arrival times of Alice's and Bob's signal optical pulses with his SNSPD and
calculates the difference. A programmable delay chip with 10 ps timing resolution and synchronization signal delay method are exploited to compensate the difference of time. The chromatic dispersions of
404 km of ultralow-loss optical fibre and 311 km of standard optical fibre are both less than 30 ps, which is much smaller than the signal pulse width of 2.5 ns.
For circumventing the time drift of long distance fibre, we calibrate the time synchronization of systems every 30 minutes.

To guarantee the indistinguishability of spectrum, we elaborately select two semiconductor laser diodes with nearly identical system parameters, such as central wavelength, spectrum shape and pulse shape. Exploiting the Fabry-Perot interferometer with a resolution of 0.06 pm, the FWHM of laser pulse spectrum is 3.2 pm.
By using optical spectrum analyzer and the temperature controller inside the lasers, the central wavelength of laser can be feedback controlled.
The central wavelength difference of two lasers is less than 0.3 pm for each spectrum calibration.
The laser central wavelength will almost not change within an hour if it steadily emits optical pulses.

A fast axis blocked polarization maintaining BS is utilized to maintain polarization consistency and good interference. Before the two ways of signal optical pulses enter the BS to interfere, we add an electric polarization controller and a polarization beam splitter to adjust the polarization, respectively. Exploiting the extra SNSPD with 30 dark counts per second, we can maintain high visibility real time feedback even in the experiment with 404 km of ultralow-loss optical fibre, because the effective count per second of SNSPD is still about 3900.

A phase stabilization system is improved to share the stabilized phase reference frame between Alice and Bob\cite{Liu:MIQKDexp:2013}.
A phase feedback laser generates pulses with the width of 1 ns  and the central wavelength of 1550.12 nm to go through the AMZIs of Alice and Bob. A stabilized and good interference of the feedback optical pulse represents the low misalignment of phase reference frames, which is feedback controlled by a phase shifter and avalanche photodiode single-photon detector. To  decrease the error of X basis, the central wavelength of the phase feedback laser is also adjusted by the optical spectrum analyzer and temperature controller every 4 fours and the AMZIs of Alice and Bob are carefully developed with a high visibility of 200:1.

\end{methods}

We thank Y. Tang, X. Xie, M. Jiang for valuable discussions. This work was supported by the National Fundamental Research Program (under 2013CB336800), the National Natural Science Foundation of China, the Chinese Academy of Science, the 10000-Plan of Shandong Province, the Science Fund of Anhui Province for Outstanding Youth, the National High-Tech Program of China (under Grant No. 2011AA010800 and No. 2011AA010803) and the QuantumCTek Co., Ltd.

%
%

\newpage

\newpage

\newpage

\newpage

\begin{table}
\caption{\label{tab:Para} \textbf{Optimized intensities ($\mu_\alpha$) and probabilities ($p_\alpha$) for each distance in our experimental setup.}}
\begin{tabular}{c|cccccc}
  \rm{Distance} & 102km & 155km & 207km & 259km & 311km & 404km  \\
  \hline
  $\mu_z$ & 0.891 & 0.864 & 0.757 & 0.677 & 0.453 & 0.413 \\
  $\mu_y$ & 0.189 & 0.191 & 0.203 & 0.267 & 0.363 & 0.302 \\
  $\mu_x$ & 0.049 & 0.058 & 0.059 & 0.064 & 0.083 & 0.073 \\
  $p_z$ & 0.827 & 0.789 & 0.731 & 0.509 & 0.409 & 0.315 \\
  $p_y$ & 0.025 & 0.038 & 0.042 & 0.068 & 0.101 & 0.110 \\
  $p_x$ & 0.128 & 0.154 & 0.201 & 0.388 & 0.439 & 0.529 \\
  $N_{t}$ &$2.05\times10^{12}$ & $2.03\times10^{12}$ & $3.61\times10^{13}$ & $3.55\times10^{13}$ & $9.09\times10^{13}$ & $6.04\times10^{14}$ \\
  \hline
\end{tabular}
\end{table}

\end{document}